\documentclass[a4paper,12pt]{article}
\usepackage{amsmath}
\usepackage{amsfonts}
\usepackage{amssymb}
\usepackage{graphicx}
\usepackage{time}
\usepackage{relsize}
\usepackage[utf8]{inputenc}
\usepackage{lipsum}
\usepackage{setspace}

\usepackage[left=1.25cm,right=1.25cm,top=2.2cm,bottom=2.5cm]{geometry}
\usepackage[colorlinks]{hyperref}
\usepackage[usenames,dvipsnames]{color}
\hypersetup{
    breaklinks=true,
    pdfstartview={FitH},    
    colorlinks=true,       
    linkcolor=magenta,          
    citecolor=Cyan,        
    filecolor=magenta,      
    urlcolor=blue,           
    anchorcolor=green,      
    linktocpage=true
}
\makeatletter
\def\vhrulefill#1{\leavevmode\leaders\hrule\@height#1\hfill \kern\z@}
\makeatother

\newcommand{\nc}{\newcommand}
\nc{\ba}{\begin{eqnarray}}
    \nc{\ea}{\end{eqnarray}}
\newcommand\be{\begin{equation}}
    \newcommand\ee{\end{equation}}

\def\thefootnote{\arabic{footnote}}
\begin{document}

    \vspace{5mm}
    \vspace{0.5cm}
    \begin{center}

        \def\thefootnote{\fnsymbol{footnote}}

        {\large \bf Superradiant energy extraction from rotating hairy Horndeski black holes }
        \\ \vspace{0.2cm} \vhrulefill{1.6pt}

        \vspace{0.5cm}
        {\bf Sohan Kumar Jha$^a$}, \footnote{\url{sohan00slg@gmail.com}}
        {\bf Mohsen Khodadi$^{b,c}$}, \footnote{\url{m.khodadi@hafez.shirazu.ac.ir}}
        {\bf Anisur Rahaman$^d$}, \footnote{\url{anisur@associates.iucaa.ac.in}}
        {\bf Ahmad Sheykhi$^{b,c}$} \footnote{\url{asheykhi@shirazu.ac.ir}}
        \\[0.4cm]

        { $^a$\textit{Chandernagore College, Chandernagore, Hooghly, West Bengal, India}}\\

        { $^b$\textit{Department of Physics, College of Sciences, Shiraz University, Shiraz 71454, Iran}}\\

        {$^c$\textit{Biruni Observatory, College of Sciences, Shiraz University, Shiraz 71454, Iran}}\\

        {$^d$\textit{Durgapur Government College, Durgapur, Burdwan - 713214, West Bengal, India}
        }\\

        \date{\today}

    \end{center}

    \vspace{.01cm}

    \begin{abstract}
Adopting the manifest of low-frequency and low mass for the scalar
perturbation, we perform a semi-classical analysis of the
superradiance phenomenon for a rotating hairy Horndeski black hole
(BH). For the spacetime under study enriched by the hairy
Horndeski  parameter $h$, in addition to the mass $M$ and spin
$a$, we compute the amplification factor of scalar wave scattering
indicating the energy extraction from the BH. We find that due to
the addition of the hairy parameter $h$ in the geometry, the
superradiance scattering and its frequency range enhance compared
to the Kerr BH. This implies that Horndeski's gravity belongs to
those alternative theories of gravity that make the amplification
factor larger than the Kerr BH so that the energy extraction in
its framework is more efficient than general relativity.
Calculating the outgoing energy flux measured by an observer at
infinity verifies the role of the hairy parameter $h$ in the
increase in energy extraction efficiency from the rotating BH. By
implementing the BH bomb mechanism, we present an analysis of the
superradiant instability
of the underlying BH spacetime against massive scalar fields. Our analysis indicates that the hairy Horndeski parameter leaves no imprint on the standard superradiant instability regime.\\\\

\end{abstract}

    \newpage
    \vspace{0.5cm} \vhrulefill{1.2pt}
    \tableofcontents
    \vspace{0.5cm} \vhrulefill{1.2pt}
    \section{Introduction}
The exciting idea of extracting energy from a black hole (BH)
through the amplification process dates back to the pioneering
work of Penrose \cite{Penrose:1969pc,Penrose:1971uk}, who proposed
energy extraction for a particle falling and decaying into the
ergoregion of a Kerr BH \footnote{It can be interesting to point
out that energy extraction via the Penrose process is not
affordable astrophysical and instead by regarding the magnetic
field around BH its efficiency improves
\cite{Wagh:1985vuj,Dadish:1986,Wagh:1989zqa}. Recently Ref.
\cite{Comisso:2020ykg}, has shown that the magnetic field
enclosing rotating BH, via a theoretical mechanism known as
magnetic reconnection, has the potential for energy extraction
with more efficiency, see also follow-up studies
\cite{Khodadi:2022dff,Carleo:2022qlv,Wei:2022jbi}.}.  Indeed,
ergoregion is a region in which a timelike captured particle can
have negative energy, as perceived by an observer at infinity.
According to the energy conservation law, the swallowing of the
particles with negative energy by BH means extracting energy from
the BH. This process also can be generalized to wave scattering
off BH. Using the massless scalar field, Misner derived essential
inequality $\omega<m\Omega$ between the frequency of the incitant
wave $\omega$ and the rotational frequency of the BH $\Omega$
\cite{Misner:1969hg}. This, in essence, is known as superradiant
scattering i.e., the amplification of waves when scattering off a
dissipative rotating body. By analyzing a dissipative system such
as an absorber rotating cylinder subject to scattering of waves,
Zeldovich derived successfully the mentioned condition of
rotational superradiance ($\omega<m\Omega$)
\cite{Zel:1971,Zel:1972}. Indeed, it is well-known that an
incident wave in case of scattering off any dissipative object is
prone to experiencing superradiance. In this regard, Teukolsky
\cite{Teukolsky:1973ha} has shown that the Misner/Zeldovich
amplification process occurs also for other bosonic fields
(electromagnetic and gravitational waves) provided that satisfies
the inequality above. Further, by analyzing BH superradiance from
the perspective of thermodynamics, Bekenstein derived
\cite{Bekenstein:1973mi} results in agreement with previous ones
(see also \cite{Bekenstein:1998nt}). The milestone of Bekenstein's
study was the discovery of a fundamental origin for the energy
extraction via connecting it to Hawking's theorem, expressing that
the surface area of a classical BH cannot decrease
\cite{Hawking:1971tu}. Historically, these seminal papers on the
BH superradiance were the first steps that later resulted in
discovering BH evaporation by Hawking \footnote{For this reason
that by taking the quantum effects into account, the rotational
superradiance would become a spontaneous process and BH would slow
down by spontaneous emission of photons satisfying
$\omega<m\Omega$.} \cite{Hawking:1975vcx}. This along with some of
the newer studies on this subject such as
\cite{Richartz:2009mi,Cardoso:2012zn}, may support the idea that
superradiant scattering strongly depends on the boundary
conditions at the horizon. This implies that superradiant
scattering is nothing but a boundary condition problem. However,
several evidences reveal that ergoregion is an essential component
for occurring  superradiance phenomena, since it provides required
friction as a form of dissipation
\cite{Eskin:2015ssa,Vicente:2018mxl} (see also
\cite{Brito:2015oca} for review). Based on this, horizonless
compact objects such as stars  are also prone to trigger
superradiance \cite{Richartz:2013unq}. Indeed, due to the probable
existence  of viscous matter, stars can be regarded as a model
that may provide requisite dissipation for superradiance phenomena
to occur \cite{Cardoso:2015zqa} (see also
\cite{Glampedakis:2013jya}).

Apart from the fact that superradiant amplification can be
utilized to explain energy extraction from a BH, it may cause
several important instabilities in BH spacetimes, as well. If
these instabilities occur, due to rotational energy extraction,
the spin of BH slows down and can lead to a hairy BH
solution. Therefore, these instabilities may open a new window to
investigate the no-hair theorem for any new BH solution
\cite{Rahmani:2020vvv}. The idea behind superradiant instability
is not complicated and comes from the fact that confinement of the
system subject to perturbation causes unstable environment
against superradiant modes. By amplifying any incoming pulse near
the ergoregion by superradiance, and then confining the pulse via
a perfect reflecting surface at some distance, the amplitude of
the pulse exponentially increases through numerous interactions.
It means making instability in the background subject to some
perturbation. Commonly, a reflecting surface is supplied in two
different ways: natural and artificial. The former can be
achieved if there is any provision for  Anti-De-Sitter (AdS)
spacetime\footnote{In Ref. \cite{Konoplya:2011it} it has shown
that the asymptotically Godel spacetime is also similar to AdS
which can play the natural role of a time-like boundary. Note that
no time-like particle can reach spatial infinity, therefore any
background similar to AdS can be considered as a confining system.
}
\cite{Cardoso:2006wa,Li:2012rx,Aliev:2015wla,Green:2015kur,Rahmani:2020wlq}
and the massive scalar field
\cite{Dolan:2007mj,Hod:2012zza,Hod:2016kpm,Huang:2018qdl,Xu:2020fgq,Vieira:2021nha,Myung:2022yuo},
while the latter can be reached by placing a mirror-like surface
around BH or confining the BH into a box with Dirichlet boundary
conditions
\cite{Witek:2010qc,Herdeiro:2013pia,Degollado:2013bha,Hod:2013fvl}.
Press and Teukolsky \cite{Press:1972zz} explained the mechanism of
the instability in this way that the initial fluctuation arising
from superradiant modes grows exponentially, leading to an
ever-increasing field density and pressure inside the confinement
region such that finally disrupts the confining surface, resulting
in an explosion. This system is known as the BH bomb, see Ref.
\cite{Cardoso:2004nk} for more details. Another motivation for
studying superradiant instability of BH comes from the fact that
constraining the mass of ultra-light degrees of freedom may shed
light on the dark matter puzzle
\cite{Hannuksela:2018izj,Brito:2020lup,Ng:2020ruv,Yuan:2022bem}.

In recent years, this subject has received a considerable
attention from different perspectives, including astrophysics,
higher dimension spacetimes, and also alternative theories of
general relativity. The importance of superradiant instability in
the framework of astrophysics originates from the fact that its
development due to the extraction of energy and angular momentum
from the BH results in the formation of a non-spherical bosonic
cloud near the BH and subsequently gravitational wave emission
\cite{Witek:2012tr,Arvanitaki:2014wva,East:2018glu} (see also
recent papers \cite{Khodadi:2022dyi,Siemonsen:2022yyf}). The
application of the gravitational wave emitted by the cloud is that
creating specific quasi-monochromatic signals would address the
existence of ultra-light bosons. In other words, the superradiant
instabilities allow us to use astrophysical BHs as effective
detectors \cite{Brito:2014wla} to look for new particles
\cite{Blas:2022fvy}. It was argued that the superradiant
instability at interplay with the BH shadow potentially can be
used to constrain the ultra-light bosons candidates
\cite{Roy:2019esk,Creci:2020mfg,Roy:2021uye,Chen:2022nbb}. Other studies
\cite{Cunha:2019ikd,Saha:2022hcd,Shakeri:2022usk,Khodadi:2022ulo,Chen:2022kzv}
propose the possibility of probing ultra-light bosons using the
superradiant clouds around the supermassive BHs recorded recently
by the Event Horizon Telescope. The investigations on the
superradiant phenomenon for higher-dimensional models are
well-motivated since according to these models in particle
accelerators such as LHC (Large Hadron Collider) there is a chance
to produce micro BHs \cite{Ida:2005ax}. By serving the scalar and
vector perturbations, the efficiency of superradiant amplification
as well as its instabilities for higher dimensional rotating BHs
have been addressed in
\cite{Jung:2005pk,Jung:2005cn,Cardoso:2005vk,Creek:2007sy,Creek:2007pw,
Kodama:2007sf,Casals:2008pq}.
Similar investigations have been performed for non-rotating
charged BHs \cite{Wang:2014eha,Huang:2016zoz,Destounis:2019hca}.

Another domain of interest for studying the superradiance
phenomena of BHs is modified gravity, which we shall address by
utilizing one of the well-known models beyond Einstein's gravity.
In other words, BH superradiance is not a prerogative of general
relativity, rather any relativistic gravitational theory that
admits BH solutions,
 is prone potentially to it. Despite the agreement of the recent
gravity experiments in the strong field regime
\cite{LIGOScientific:2016aoc,EventHorizonTelescope:2019dse,
EventHorizonTelescope:2022xnr}
with the standard Kerr BH, due to the statistical error in current
observations, the possibility of admitting the Kerr BH solutions
modified by some alternative theories of gravity is still not
ruled out. In general, the details of the superradiant energy
extraction are affected by two factors: BH geometry and the wave
dynamics in the alternative theory of gravity. The motivation for
the study of superradiance within the modified gravities is
twofold. First, Einstein's gravity, despite all its admirable
achievements, cannot be a reliable theory at all scales due to
some shortcomings, and it is expected that it needs some
modifications \cite{Capozziello:2011et}. Second, finding the
imprint of corrections imposed to the standard model of gravity on
the superradiance efficiency and its instabilities is also
potentially interesting. In recent years, we are seeing a variety
of research on this topic (see e.g.,
\cite{Cardoso:2013opa,Cardoso:2013fwa,Kolyvaris:2018zxl,
Frolov:2018bak,Wondrak:2018fza,Khodadi:2020cht,Franzin:2021kvj,
Khodadi:2021owg,Cuadros-Melgar:2021sjy,Khodadi:2021mct,
Jha:2022ewi,Siqueira:2022tbc,Jha:2022nzd,Yang:2022uze})
which compared to standard general relativity enhances or
subsidies. Such investigations allow us to separate those modified
gravities which are in favor of superradiance from those that are
not. Despite these studies, there is still empty room for some
classes of BH solutions. In the present work, we focus on hairy
rotating BHs that arise from Horndeski theory of gravity.

Introducing the scalar field is one of the well-known ways to
extend gravity to overcome cosmological issues, including dark
energy, dark matter, and the evolution of the universe in early
and late-time epochs. There are some modified gravity theories
that are mathematically equivalent to a gravitational theory with
degrees of freedom containing the metric $g_{\mu\nu}$ and one or
more scalar fields $\phi$
\cite{Capozziello:2005mj,Sotiriou:2006hs,Capone:2009xk,Ntahompagaze:2017dla}.
The scalar-tensor theories are likely the simplest, most
consistent, and nontrivial extensions of general relativity
\cite{Damour:1992we}. One of the most famous four-dimensional
scalar-tensor theory is the Horndeski gravity proposed in 1970
\cite{Horndeski:1974wa}. In the light of some leading research
\cite{DeFelice:2010nf,Deffayet:2011gz}, it was argued that Horndeski gravity is equivalent to the Galilean theories which, in essence, are the scalar-tensor theories with Galilean symmetry in flat spacetime
\cite{Kobayashi:2011nu,Kobayashi:2019hrl}. Indeed, due to
propagating one scalar degree of freedom by general second-order
field equations in the Horndeski gravity, it is free of
Ostrogradski instabilities.
In Refs. \cite{Lombriser:2015sxa,Bettoni:2016mij}, authors have shown that gravitational waves are an efficient and robust tool for distinguishing the models of the Horndeski theory describing the cosmic accelerating expansion. One of the freedoms that appear in Horndeski's theories is that, unlike the standard general relativity, there is no requirement that gravitational waves (tensor speed $c_T$) travel at the speed of light in the vacuum i.e., $c_T=c$.
Apart from the key physical reason giving rise to such anomalous propagation, gravitational waves when analyzed in the framework of modified gravity no longer travel on null geodesics of the background metric as photons do \cite{Perenon:2019qmd}. The release of the observational data of GW170817 and whose optical counterpart GRB170817A, have placed very tight constraints on the deviation from the speed of light \cite{LIGOScientific:2017zic}. In general, the results of GW170817 and GRB170817A indicate that the deviation of the tensor speed (gravitational waves) from the speed of light is no more than one part in $10^{15}$ i.e., $\mid c_T-1\mid\lesssim 10^{15}$. Thanks to this tight constraint, it is possible to test the validity of Horndeski's descriptions of late-times cosmological evolution (e.g., see Refs. \cite{Creminelli:2017sry,Ezquiaga:2017ekz,Copeland:2018yuh}).
Cosmological consequences of this
theory such as alleviating the cosmological constant problem
\cite{Charmousis:2011bf} and other interesting features
\cite{Nicolis:2008in,Maselli:2016gxk,Babichev:2016rlq,Petronikolou:2021shp,Ripley:2022cdh,Kubota:2022lbn}
have been studied. While, observational constraints  on the
parameters of Horndeski theories have been carried out in
\cite{Bellini:2015xja,Bhattacharya:2016naa,Kreisch:2017uet,Hou:2017cjy,SpurioMancini:2019rxy,Allahyari:2020jkn},
thermodynamics of BH solutions of this theory were explored in
\cite{Feng:2015oea,Feng:2015wvb,Miao:2016aol,Hajian:2020dcq,Sang:2021rla,Giusti:2021sku}.
Taking hairy BH solution \footnote{One can imagine the Kerr
hypothesis as a consequence of the no-hair theorem, meaning that
the endpoint of any gravitational collapse will be nothing but a
Kerr BH. However, this is different in the framework of modified
theories and it is expected to there exist classes of these
theories that predict the hairy BH solutions. Note that the
existence of hairy BH solutions always is not meaning that there
is an additional new conserved charge (for a comprehensive
overview refer to Refs. \cite{Herdeiro:2015waa,Cardoso:2016ryw}).
Already several static and  spherically symmetric hairy BHs in the
framework of scalar-tensor theories were obtained which as the
simplest case can mention those solutions with a radial dependency
scalar field
\cite{Sotiriou:2013qea,Sotiriou:2014pfa,Benkel:2016rlz}. Apart
from the theoretical aspects, some of these hairy solutions have
also been subjected to observations
\cite{Khodadi:2020jij,Khodadi:2021gbc}. }
\cite{Bergliaffa:2021diw} derived from the quartic scalar field
version of the Horndeski gravity, the role of adding a hairy
parameter on the strong gravitational effects such as lensing
\cite{Kumar:2021cyl}, deflection of light \cite{Walia:2021emv},
and BH shadow \cite{Afrin:2021wlj} have been explored.

In line with mentioned above, in the present work, we are going to investigate the influence of the hairy parameter, admitted by the quartic scalar field Horndeski gravity, on massive scalar field superradiant amplification, and the relevant stability linked with it. This study is well-motivated in the sense that it allows us to address the phenomenological imprint induced by the correction in the geometric structure of spacetime.
Throughout this paper, we use the signature convention $(-,+,+,+)$
and work in the units where $c = 1=\kappa = 8\pi G_N$.

This paper is structured as follows. In Sec. \ref{Horn}, by serving
 the quartic Horndeski scalar field model, we present the rotating
 counterpart for the spherically symmetry hairy BH solution \cite{Bergliaffa:2021diw},
  as already was released in \cite{Walia:2021emv}. Sec. \ref{super} consists
of three parts. We first address the conditions of massive scalar
superradiant scattering, and then by serving a semi-classical
technique, we compute analytically the amplification factor to
look for the role of hairy Horndeski in strengthening/weakening of
scalar wave. In the third part of Sec. \ref{super}, we investigate
the energy extraction from the rotating Horndeski BH. In
Sec. \ref{Ins}, by analyzing the effective potential in the
framework of the BH bomb mechanism, we discuss the superradiant
instability of the dynamics of the massive scalar fields. We
finish with closing remarks in Sec. \ref{conc}.
\section{Rotating Horndeski black hole metric}\label{Horn}
The action of Horndeski gravity consists of the metric
$g_{\mu\nu}$ and the scalar field $\phi$ include four arbitrary
functions $Q_{i=2,..5}(\chi)$ of kinetic term
$\chi=-1/2\partial_\mu\phi\partial^\mu\phi$
\cite{Babichev:2017guv}. Adopting the special case in which $Q_
5=0$, the quartic action of Horndeski gravity can be expressed as
    \begin{eqnarray}\label{action}
        S=\int d^4x\sqrt{-g}\Bigg{\{}Q_2(\chi)+Q_3(\chi)\Box\phi+Q_4(\chi)R+Q_{4,\chi}\bigg((\Box\phi)^2
        -(\triangledown^\mu\triangledown^\nu\phi)(\triangledown_\mu\triangledown_\nu\phi)\bigg)\Bigg{\}},
    \end{eqnarray}
where $g$, $R$, $\Box$ and $\triangledown$ denote the determinant
of the metric tensor, the Ricci scalar, the d’Alembert operator
and the covariant derivative, respectively.

The spherically symmetric hairy
 Horndeski BH spacetime has the following line elements
\cite{Bergliaffa:2021diw}
 \begin{eqnarray}\label{nr}
    ds^2=-A(r)dt^2+B^{-1}(r)dr^2+C(r)\big( d\theta^2 +\sin^2\theta d\varphi^2\big)~,~~~C(r)=r^2
 \end{eqnarray}
By varying the action (\ref{action}) with respect to $\phi_{,\mu}$, and $g^{\mu\nu}$, we  obtain
\begin{eqnarray}
    \label{eQ_2}
    j^\nu=&-Q_2,_\chi \phi^{, \nu}-Q_3,_\chi (\phi^{, \nu}\square\phi+\chi^{, \nu})~
    -Q_4,_\chi (\phi^{, \nu}R-2R^{\nu\sigma}\phi,_\sigma)~~~~~~~~~~~~~~~~~~~~~~~~~~~\\
    \nonumber
&   -Q_4,_\chi,_\chi\big(\phi^{, \nu}[(\square \phi)^2
    -(\nabla_\alpha\nabla_\beta\phi)(\nabla^\alpha\nabla^\beta\phi)]
        +2(\chi^{, \nu}\square\phi-\chi,_\mu\nabla^{\mu}\nabla^{\nu}\phi)\big),
\end{eqnarray}
and
\begin{eqnarray}
    \label{eq3}
    Q_4 G_{\mu\nu}= T_{\mu\nu},
\end{eqnarray}
where are respectively four-current and the field equations with
\begin{eqnarray}
    \label{eq4}
    \nonumber
    T_{\mu\nu}=&&\frac{1}{2}(Q_2,_\chi \phi ,_\mu \phi ,_\nu+Q_2 g_{\mu\nu})
    +\frac{1}{2}Q_3,_\chi(\phi ,_\mu \phi ,_\nu\square\phi
    -g_{\mu\nu} \chi,_\alpha \phi^{, \alpha}+\chi,_\mu \phi ,_\nu
    +\chi,_\nu \phi ,_\mu)~~\\
&&  - Q_4,_\chi\Big(\frac{1}{2}g_{\mu\nu}[(\square\phi)^2
    -(\nabla_\alpha\nabla_\beta\phi)(\nabla^\alpha\nabla^\beta\phi)-2R_{\sigma\gamma}\phi^{,\sigma}\phi^{,\gamma}]
    -\nabla_\mu\nabla_\nu \phi \square\phi\\ \nonumber
    &&
    +\nabla_\gamma\nabla_\mu \phi \nabla^\gamma \nabla_\nu \phi-\frac{1}{2}\phi ,_\mu \phi ,_\nu R
    +R_{\sigma\mu}\phi^{,\sigma}\phi,_{\nu}
    +R_{\sigma\nu}\phi^{,\sigma}\phi,_{\mu}+R_{\sigma\nu\gamma\mu} \phi^{,\sigma}\phi^{,\gamma}
    \Big)~~~~~~~~~~~~~~\\
    \nonumber
&&  -Q_4,_\chi,_\chi \Big(g_{\mu\nu}(\chi,_{\alpha}\phi^{,\alpha}\square \phi+\chi_{,\alpha} \chi^{, \alpha})+\frac{1}{2}\phi ,_\mu \phi ,_\nu
    (\nabla_\alpha\nabla_\beta\phi\nabla^\alpha\nabla^\beta\phi
    -(\square\phi)^2)\\ \nonumber
&&- \chi,_\mu \chi,_\nu
    - \square\phi( \chi,_\mu \phi,_\nu
    + \chi,_\nu \phi,_\mu)  - \chi,_\gamma[\phi^{,\gamma}\nabla_\mu\nabla_\nu\phi-(\nabla^\gamma\nabla_\mu\phi)\phi,_{\nu}
    -(\nabla^\gamma\nabla_\nu\phi)\phi,_{\mu}]
    \Big).
    \end{eqnarray}
To have spherically symmetric spacetime, as addressed by the metric (\ref{nr}), it is essential one set a scalar field $\phi\equiv\phi(r)$. Without going into the details of \cite{Bergliaffa:2021diw}, by introducing simple forms for $Q_{2}= \alpha_{22} (-\chi)^{3/2}$, $Q_{3}=0$ and $Q_{4}=\kappa^{-2} +\alpha_{42} (-\chi)^{1/2}$ \footnote{It means that the Horndeski model under our attention throughout the paper just includes $Q_2$ and $Q_4$. Namely it belong a subclass of Horndeski theories that at same time has both proprieties: shift symmetric (i.e., symmetric under $\phi\rightarrow \phi+\mbox{constant}$) and reflection symmetric (i.e., symmetric under $\phi\rightarrow-\phi$).
The presence of $Q_2$ is vital for justify the cosmic accelerating  expansion and the gravitational wave propagation simultaneously. More exactly, constraining the Horndeski theory via simultaneous confronting with GW170817 and GRB170817A \cite{LIGOScientific:2017zic}, explicitly exclude the subclass models that contain $Q_4$ without $Q_2$ \cite{Arai:2017hxj}.}, 
and satisfying the conditions: a vanishing radial four-current at infinity $j^{r}=0$,
finiteness of the energy of $\phi$ i.e., $ \int_V \sqrt{-g}\, T^{0}_{0} \,d^{3}x$, and utilizing the field equations (\ref{eq3}), the metric components of (\ref{nr}) and the derivative of the scalar field background take the following forms, respectively
\begin{align}
    \label{metric1}
    A(r)&=B(r)=1 -\frac{2M}{r} + \frac{h}{r}\ln\left({\frac{r}{2M}}\right),\\
   \phi'(r)&=\pm\frac{2}{r}\sqrt{\frac{-\alpha_{42}}{3 A(r) \alpha_{22}}}~  \label{phi}.
\end{align}

If $h\rightarrow 0$ (with the mass dimension $[h]=M$), the Schwarzschild solution is recovered. It, in essence,
related to parameters $\alpha_{22}$ and $\alpha_{42}$ respectively in the functions $Q_2$, and $Q_4$ in the action (\ref{action}). More exactly, it reads as $h=(2/3)^{3/2}\kappa^2 \alpha_{42}\sqrt{-\alpha_{42}/\alpha_{22}}$ \cite{Bergliaffa:2021diw}. 
Recently, the range of the hairy Horndeski parameter $h$ in the scale of the supermassive black hole horizon located in the center of our galaxy has been scanned \cite{Vagnozzi:2022moj}.
As it is clear from (\ref{phi}) the derivative of the scalar field is finite and real if $A(r)>0$ (i.e., everywhere outside of the black hole) and $\alpha_{42}/\alpha_{22}<0$. Namely, the current $j^{r}$ (as only non-zero component of four-current $j^{\mu}$), is not finite exactly on the horizon, but in the vicinity of it ($A(r)\rightarrow0^+$) the behavior of both $\phi' (r)$, and $j^{r}$ are regular. Providing a more details discussion on this can be helpful to understand such a behavior of the derivative of the $\phi$ on the horizon. For the gravity-galileon field coupled systems, it is shown in Ref. \cite{Hui:2012qt} by Hui and Nicolis that a no-hair result of black holes hold under some assumptions: (a) asymptotic flatness,
	(b) vanishing derivative of the scalar field at infinity,
	(c) the finiteness of norm of the Noether four-current $j_{\mu}j^{\mu}$ down to the horizon, (d) the presence of canonical kinetic term $\chi$ in the action, (e) the $\chi$-derivatives of $Q_{2...5}$ contain only positive or zero powers of $\chi$. In other words, bypassing one or more of these assumptions in the no-go theorem above may result in a hairy black hole solution.  In \cite{Hui:2012qt} have shown that satisfying the assumption (c) results in a time-independent but constant profile for scalar field i.e. $\phi(r)=\mbox{constant}$. The hairy black hole solution in \cite{Bergliaffa:2021diw} that we are interested in is, in essence,  obtained from leaving the assumptions (c) and/or (e), along with taking a time-independent profile but nonconstant ($r$-dependent) into account of the scalar field. Indeed, assumption (c) is replaced with the finiteness of the Noether four-current $j^{\mu}$ at infinity, meaning that the energy of the scalar field in a volume outside the event horizon, is finite.
Note that the derivative of scalar field (\ref{phi}) is direct result of imposing condition $j^{r}=0$ (as $r\longrightarrow \infty$ i.e., break of assumption (c)), which comes from (\ref{eQ_2})
\begin{eqnarray}
    \label{e2}
    j^r=-Q_2,_\chi \phi'-\frac{2Q_4,_\chi,_\chi}{r^2} \phi'^3.
\end{eqnarray} Because it is expected that the profile of the underlying time-independent scalar field obeys the metric symmetries, only the non-zero component of the four-current is radial $j^r$. The divergence of the derivative of scalar field profile (\ref{phi}) on the horizon merely does not restrict to the hairy BH solution under our attention rather for a wide class of Horndeski gravity theories, one can find slowly rotating black hole solutions with such behavior, see Ref. \cite{Maselli:2015yva}.
In this direction also by setting coupling functions $Q_{2}\propto \chi$, $Q_{3}=0$ and $Q_{4}\propto (-\chi)^{1/2}$ in the quartic Horndeski theory, authors in \cite{Babichev:2017guv} constructed spherically symmetric and static BHs with the nonconstant profile for the scalar field that its derivative diverges on the horizon. As another example in Ref. \cite{Babichev:2017guv} one can mention also a special subclass of Horndeski theories violating assumption (d) and admitting the standard Kerr metric with a non-trivial profile for the scalar field that its derivative is regular everywhere outside of the Kerr BH except at the event horizon.
Here it is required to comment on  the irregular behavior of $\phi'(r)$ on the horizon. Although $\phi'(r)$ is not regular on the horizon, in Ref. \cite{Bergliaffa:2021diw} authors have discussed that the components of the energy-momentum tensor (\ref{eq4}) calculated using scalar field profile (\ref{phi}) in agreement with the components of the Riemann tensor, on the horizon and outside of it are finite.
In this way, the metric of background solution (\ref{metric1}), effectively lets us take a primary step in direction of finding an intuition of the footprint of the Horndeski field on the phenomenology of BHs.

Recently, in Ref. \cite{Walia:2021emv} provided the following
rotational version for the spherically symmetric spacetime (\ref{nr}) with laps function (\ref{metric1})
    \begin{eqnarray}\label{rotmetric}
        ds^2=g_{\mu\nu}dx^{\mu}dx^{\nu}&=&-\left(\frac{\tilde{\Delta}-a^2 \sin^2\theta}{\Sigma}\right)dt^2+\frac{\Sigma}{\tilde{\Delta}}dr^2-2a\sin^2\theta\left(\frac{\tilde{\Delta}-(r^2+a^2)}{\Sigma}\right) dtd\varphi+\nonumber\\&&
        \Sigma d\theta^2 +\sin^2\theta\left(\frac{(r^2+a^2)-a^2\tilde{\Delta}\sin^2\theta}{\Sigma}\right)d\varphi^2,
    \end{eqnarray}
    where
    \begin{eqnarray}
 \Sigma=r^2+a^2 \cos^2\theta,~~~~~~      \tilde{\Delta}=r^2-2Mr+a^2+h r\ln\left(\frac{r}{2M}\right).
    \end{eqnarray}

   \begin{figure}[ht!]
      \begin{tabular}{c}
           \includegraphics[width=.45\linewidth]{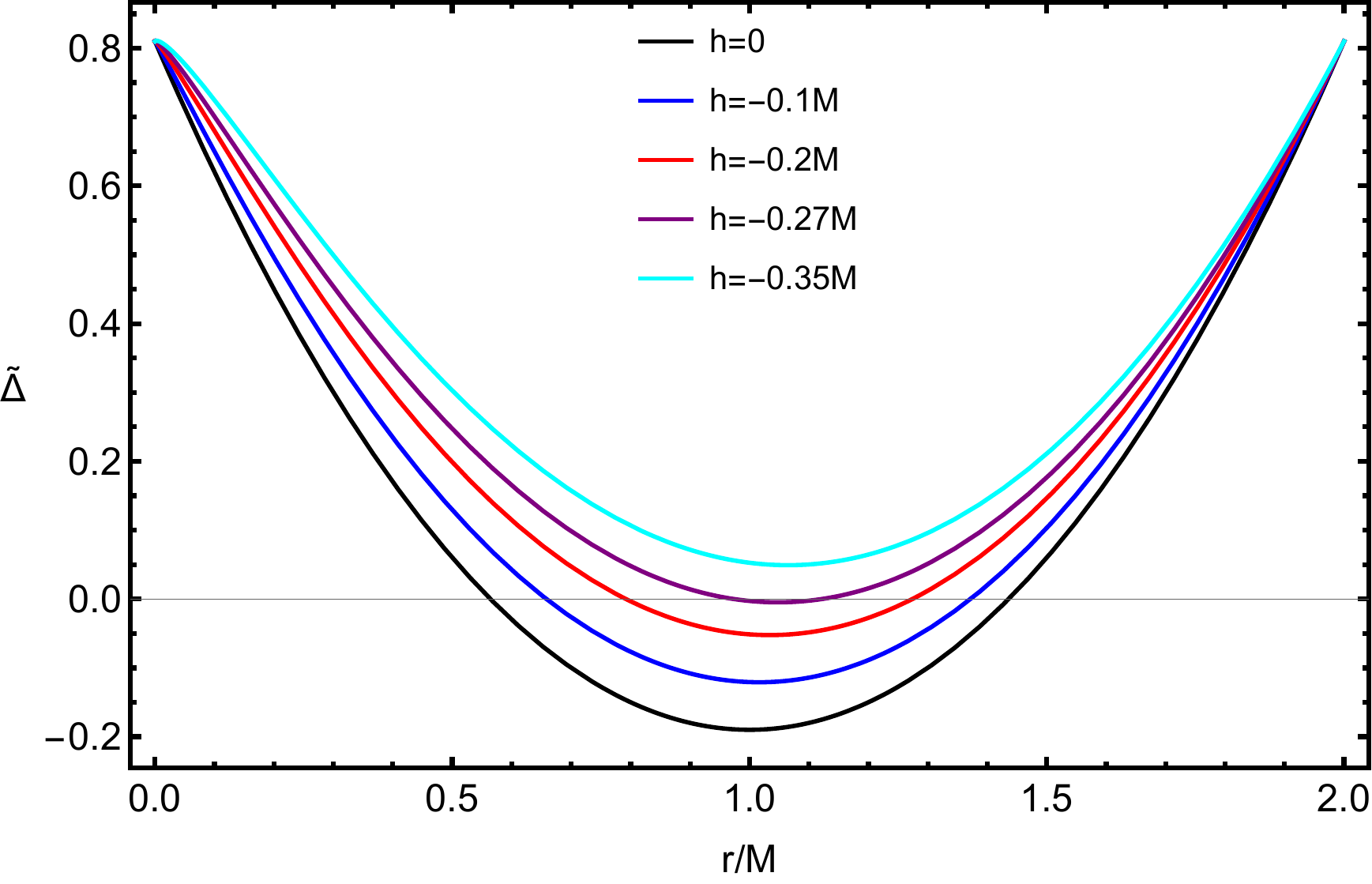}~~~~~
            \includegraphics[width=.45\linewidth]{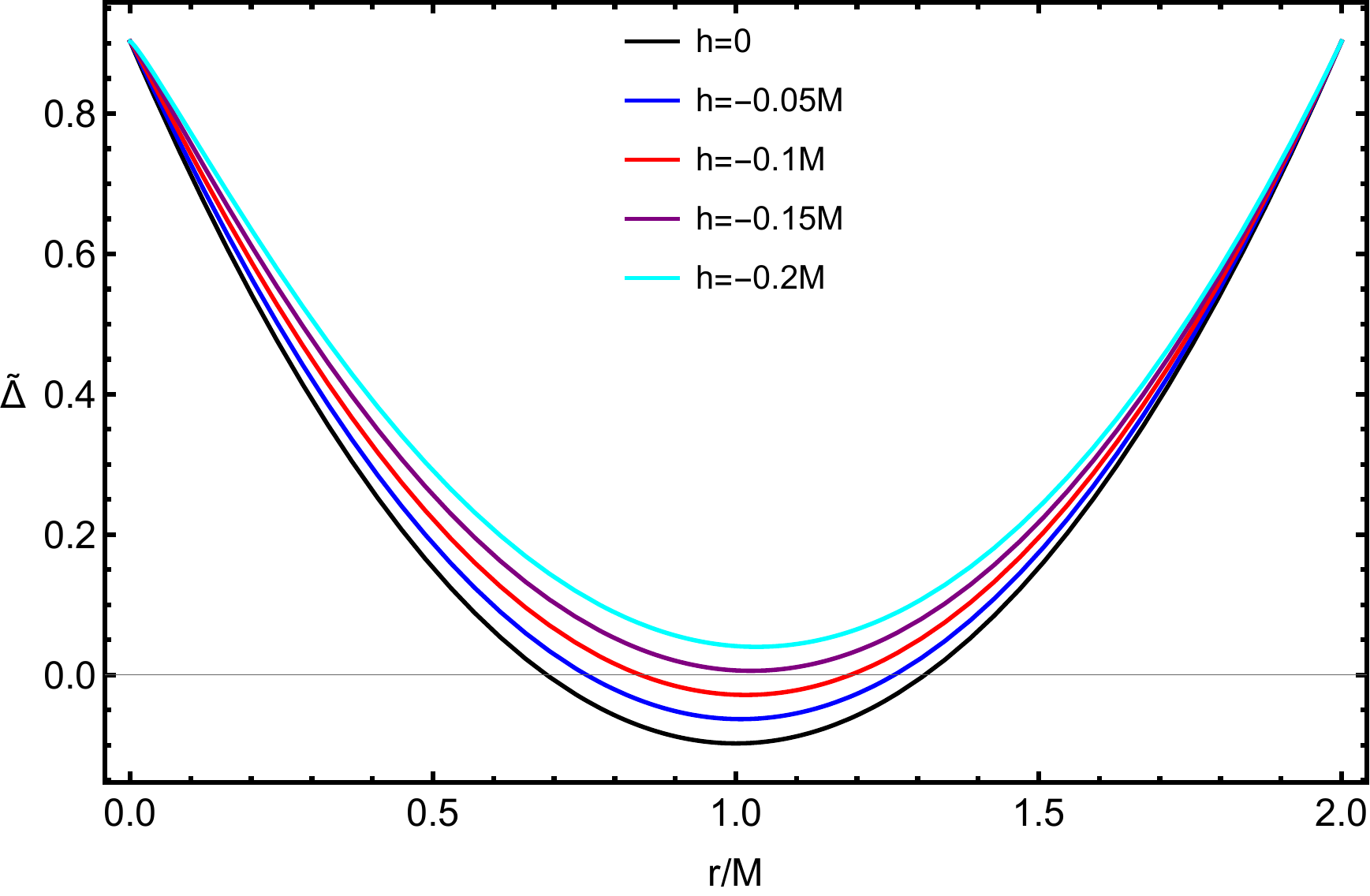}
        \end{tabular}
        \caption{\textit{The location of horizons of the rotating Horndeski BH with various values of hairy parameter $h$, and
                spin parameters: $a=0.9M$ (left panel) and $a=0.95M$ (right panel). } }\label{delta1}
    \end{figure}

The metric~(\ref{rotmetric}), in essence, has been derived using the modified Newman-Janis-algorithm (i.e., Azreg-A$\ddot{\text{i}}$nou's non-complexification procedure \cite{Azreg-Ainou:2014pra}), without showing the relevant Horndeski scalar field profile in the rotating background. As stated above, the scalar field must respect the symmetries of the metric. This lets us evaluate the regularity of the Horndeski scalar field on the horizon, infinity, and along the symmetry axis. Already Ref. \cite{Babichev:2017guv} has shown that some subclasses of Horndeski gravity admit the standard Kerr BH with a non-trivial scalar field profile as an exact solution. Despite the lack of such studies for the modified Kerr metric, it is well-known from Ref. \cite{Sotiriou:2013qea} that the scalar field profile does not affect by the rotation parameter at first order i.e., $\mathcal{O}(a^2)$. Besides, since in the linear order of rotation, the spherical symmetry still holds, thereby, the only non-zero component of the four-current for the time-independent scalar field is still radial. As a result, the expression (\ref{e2}) for the current $j^r$, and subsequently the derivative of the scalar field in (\ref{phi}), remain unchanged in slowly rotating BH solutions. However, this argument works just for moderate BHs, and doing some studies aimed at deriving the non-trivial scalar field profile for full rotating BH solutions in Horndeski's theories is essential. This deserves comprehensive and separate research in the future. In any case, the lack of an analytical expression for the profile of scalar field with axial symmetry does not prohibit the study of gravitational lensing in \cite{Walia:2021emv}, and superradiance scattering here. Since in both one, just the rotating background metric (\ref{rotmetric}) is essential, not the profile of the Horndeski scalar field. In particular, concerning the latter, we will consider a test massive scalar field $\Phi$, propagating in the background deviated from the standard Kerr, i.e.,  (\ref{rotmetric}).

In the limiting case where $h=0$,
the metric~(\ref{rotmetric}) coincides with the Kerr spacetime.
The positions of the event horizon ($r_{eh}$) and the Cauchy
horizon ($r_{ch}$) are given by the solution of the equation
$\tilde{\Delta}=0$ which can only be solved numerically due to the
presence of the logarithmic term. Outside the event horizon, there
is a region known as the ergosphere, covering the area from the
event horizon to the outer ergoregion i.e.,  the largest positive
real root of $g_{tt}=0$ (its smallest root, addresses the inner
ergoregion which is located behind the event horizon).

In Fig.  (\ref{delta1}) by fixing some different
values for the free parameter $h$, depicted the location
of horizons and ergoregion of the rotating Horndeski BH, respectively.
 One can see the displacements of the location
of horizons relative to the standard Kerr ($h=0$). We observe that for
some values of $h$, the compact object at hand is no longer BH,
 since the event horizon disappears. It also depends on the value of the spin setting so that for instance by setting $a=0.9M$, and $a=0.95M$, respectively for values beyond $h=-0.27M$, and $h=-0.15M$, we indeed deal with a rotating naked singularity. Given our interest in the BH case
\footnote{Although, the event horizon as the defining property of
BH, in essence, is not directly observable
\cite{Visser:2014zqa,Cardoso:2016rao,Cardoso:2019rvt}, nowadays
rotating BHs are widely admitted as astrophysical objects
\cite{Celotti:1999tg,Bambi:2015kza}.},  in the following, we have
to take care in setting values of the free parameter of the
model.

    \section{Scalar superradiance scattering}\label{super}
The propagation of the test scalar field $\Phi$ on a curved spacetime
is described by the following Klein-Gordon
    equation \cite{ Bezerra:2013iha, Kraniotis:2016maw}
    \begin{eqnarray}
        \left(\bigtriangledown_{\alpha}\bigtriangledown^{\alpha}+\mu_s^{2}\right)\Phi(t,r,\theta,\varphi)
        = \left[\frac{-1}{\sqrt{-g}}\partial_{\nu}\left(\sqrt{-g}g^{\mu\nu
        }\partial_{\mu}\right)+\mu_s^{2}\right]\Phi(t,r,\theta,\varphi)=0,
        \label{KG}
    \end{eqnarray}
where $\mu_s$ and $g^{\mu\nu}$, respectively, denote the mass of
the scalar field, and the inverse spacetime metric. By adopting
the standard separation of variables method, we  use the following
ansatz with the standard Boyer-Lindquist coordinates $(t, r,
\theta, \varphi)$
    \begin{eqnarray}
        \Phi(t, r, \theta, \varphi)=R_{\omega j m}(r) \Theta(\theta) e^{-i
            \omega t} e^{i m \varphi}, \quad j \geq 0, \quad-j \leq m \leq j,
        \quad \omega>0, \label{PHI}
    \end{eqnarray}
to separate the equation (\ref{KG}) into radial and angular parts.
Here $R_{\omega j m}(r)$ is the radial part of the wave
function and
    $\Theta(\theta)$ is the oblate spheroidal wave function. The
    symbols $j$ is the angular eigenfunction, $m$ is the angular quantum number, and $\omega$ is the positive frequency
    of the field  under investigation as measured by a faraway
    observer. The ansatz ({\ref{PHI}), cause the differential equation (\ref{KG})
    to yield two ordinary differential equations with the following radial and angular parts
        \begin{eqnarray}
            &&\frac{d}{d r}\big(\tilde{\Delta} \frac{d R_{\omega j m}(r)}{d
                r}\big)+\big(\frac{((r^2+a^{2}) \omega-am)^{2}}{\tilde{\Delta}}\big)R_{\omega l m}(r)
            \nonumber \\
            &&-(\mu_s^{2} r^2+j(j+1)+a^{2} \omega^{2}-2 m \omega
            a) R_{\omega l m}(r)=0, \label{RE}
        \end{eqnarray}
        and
        \begin{eqnarray}
            &&\sin \theta \frac{d}{d \theta}\left(\sin \theta \frac{d
                \Theta_{\omega j m}(\theta)}{d \theta}\right)+\left(j(j+1) \sin
            ^{2} \theta-\left(\left(a \omega \sin ^{2}
            \theta-m\right)^{2}\right)\right)\Theta_{\omega j m}(\theta)\nonumber \\
            && + a^{2} \mu_s^{2} \sin ^{2} \theta \cos ^{2} \theta~
            \Theta_{\omega j m}(\theta)=0~,
        \end{eqnarray} respectively. Given that we
        intend to study the scattering of the field $\Phi$, just Eq. (\ref{RE})
        is under our attention until the end of this paper. Following the earlier
        investigations (e.g., \cite{ Bezerra:2013iha, Kraniotis:2016maw}), we may
        find a general solution of the radial equation (\ref{RE}).  We now
        apply a Regge-Wheeler-like
        coordinate $r_{*}$ which is given by
        \begin{eqnarray}
            r_{*} \equiv \int d r \frac{r^2+a^{2}}{\tilde{\Delta}},
            \quad\left(r_{*} \rightarrow-\infty \quad \text{at event horizon},
            \quad r_{*} \rightarrow \infty \quad \text{at infinity} \right)
        \end{eqnarray}
        To have the equation into the desired shape, we consider a new
        radial function $\mathcal{S}_{\omega j
            m}\left(r_{*}\right)=\sqrt{r^2+a^{2}} R_{\omega j m}(r)$.
        A few steps of  algebra yields
        \begin{equation}
            \frac{d^{2} \mathcal{S}_{\omega l m}\left(r_{*}\right)}{d
                r_{*}^{2}}+V_{\omega j m}(r) \mathcal{S}_{\omega j
                m}\left(r_{*}\right)=0~, \label{RE1}
        \end{equation}
        where the effective potential $V_{\omega j m}(r)$ reads as
        \begin{eqnarray}\label{POT}
            V_{\omega j m}(r)&=&\left(\omega-\frac{m
                a}{r^2+a^{2}}\right)^{2}-\frac{\tilde{\Delta}}{\left(r^2
                +a^{2}\right)^{2}}\left[j(j+1)+a^{2}
            \omega^{2}-2 m a \omega+\mu_s^{2} r^2 +\sqrt{r^2+a^{2}}\frac{d}{dr}\left(\frac{r\tilde{\Delta}
            }{\left(r^2+a^{2}\right)^{\frac{3}{2}}}\right)\right].
        \nonumber \\
        \end{eqnarray}
        So, what results now is
        the study of the scattering of the scalar field $\Phi$ under the
        effective potential (\ref{POT}). For this purpose, it is usually
        studied the asymptotic behavior of the scattering potential at
        the event horizon and spatial infinity. The potential at the limit of the event horizon is
        \begin{eqnarray}
            \lim _{r \rightarrow r_{eh}} V_{\omega j
                m}(r)=\left(\omega-m \Omega_{eh}\right)^{2}
            \equiv k_{eh}^{2},~~~~\Omega_{eh}
            \equiv \frac{a}{r_{eh}^2+a^{2}},
        \end{eqnarray}
        and the same at spatial infinity gives
        \begin{equation}
            \lim _{r \rightarrow \infty} V_{\omega j m}(r)=\omega^{2}-\lim _{r
                \rightarrow \infty} \frac{\mu_s^{2} r^2
                \tilde{\Delta}}{\left(r^2+a^{2}\right)^{2}}=
            \omega^{2}
            -\mu_s^{2}\equiv k_{\infty}^{2}.
        \end{equation}
        It is important to observe that the potential shows constant behavior
        at both extremal points, though the values of the constants are different at
        the two extremal points. As we now know the asymptotic behavior of the potential
        at two extremal points, we can proceed to observe the asymptotic behavior of the radial solution.
        A few steps of algebra yield the following asymptotic solutions of the radial equation
        (\ref{RE1})
        \begin{equation}\label{AS}
            R_{\omega j m}(r) \rightarrow\left\{\begin{array}{cl}
            &    \mathcal{P}_{i n}^{eh}~\frac{ e^{-i k_{eh} r_{*}}}{\sqrt{r_{e h}^2
                        +a^{2}}}~~~~~~~~~~~~~~~~~~~ \text { for }~~~ r \rightarrow r_{e h} \\
             &   \mathcal{P}_{i n}^{\infty}~ \frac{e^{-i k_{\infty}
                        r_{*}}}{r}+\mathcal{P}_{r e f}^{\infty} ~\frac{e^{i k_{\infty}
                        r_{*}}}{r} ~~~ \text { for }~~~ r \rightarrow \infty
            \end{array}\right.
        \end{equation}
        Here, $\mathcal{P}_{in}^{eh}$ corresponds to the amplitude of the
        incoming scalar wave at the event horizon ($r_{eh}$) and
        $\mathcal{P}_{in}^{\infty}$ is the corresponding quantity of the
        incoming scalar wave at infinity $(\infty)$ whereas the amplitude
        of the reflected part of  scalar wave at infinity $(\infty)$ is
        $\mathcal{P}_{ref}^{\infty}$. Finally, by computing and equating
        the Wronskian at the event horizon ($W_{eh}$) and spatial infinity ($W_{\infty}$)
         we obtain the following relation
        \begin{equation}
            \left|\mathcal{P}_{r e f}^{\infty}\right|^{2}=\left|\mathcal{P}_{i
                n}^{\infty}\right|^{2}-\frac{k_{e
                    h}}{k_{\infty}}\left|\mathcal{P}_{i n}^{e h}\right|^{2}.
            \label{AMP}
        \end{equation}
        The featured point in the above relation is that it is free of the details
        of the potential $V_{\omega j m}(r)$ in the Schr\"{o}dinger-like differential
        equation (\ref{RE1}). The relation (\ref{AMP}) tells us that the scalar wave is
        superradiantly amplified, if $\frac{k_{e h}}{k_{\infty}}<0$ i.e., $\omega<m\Omega_{e h}$.

        \subsection{Amplification factor $Z_{jm}$ for superradiance}
        In direction of our purpose i.e., the study of the cross-section
        of the scalar field $\Phi$ scattering from the Horndeski-based rotating
        BH, we employ the
        asymptotic matching
        procedure proposed in the seminal papers \cite{Starobinsky:1973aij, STRO2}. Despite the lack of an exact solution for the singularly perturbed differential equation (\ref{RE}), still providing an approximation solution via asymptotic expansions in relevant extremal points, is possible. More precisely, in this method, one indeed finds two approximate solutions each one valid for part of the range of the independent variable so that eventually by their matching, one acquires a reliable single approximate solution. It is important to point out that matching is possible just provided that the relevant expansions have a domain of overlap, meaning that the exact solutions derived for two asymptotic regions are matched in an overlapping region.
            The mentioned procedure is semi-analytical and based on two assumptions:
          the slowly rotating $a\omega\ll1$ and the gravitational size of the BH
          is much smaller than the Compton wavelength of the scalar field $\Phi$ i.e.,
          $M\omega\ll1$ (or $\mu_sM\ll1$). By rewriting the radial equation (\ref{RE})
          in the following form
        \begin{eqnarray}\nonumber
            &&\tilde{\Delta}^{2} \frac{d^{2} R_{\omega j m}(r)}{d r^{2}}+\tilde{\Delta}
            \frac{d \tilde{\Delta}}{d r} \cdot \frac{d R_{\omega j m}(r)}{d r}\\
            &&+\left(\left(\left(r^2+a^{2}\right)
            \omega-a m\right)^{2}-\tilde{\Delta}\left(\mu_s^{2}
            r^2+j(j+1)+a^{2} \omega^{2}-2 m a \omega\right)\right)
            R_{\omega j m}(r)=0~, \label{RE2}
        \end{eqnarray} we derive separate solutions related to the two overlapping
        regions, namely
         the near-region $r-r_{eh}\ll \omega^{-1}$, and the far-region $r-r_{eh}\gg M$,
          and finally, by using the matching procedure we get a single solution.

        With the change of variable $z=\frac{r-r_{eh}}{r_{eh}-r_{ch}}$
        and taking the approximation $a \omega \ll
        1$ into account, the equation (\ref{RE2}) for the near-region
        turns into
        \begin{eqnarray}
            z^{2}(z+1)^{2} \frac{\mathrm{d}^{2} R_{\omega j m}(z)}{\mathrm{d}
                z^{2}}+z(z+1)(2 z+1) \frac{\mathrm{d} R_{\omega j
                    m}(z)}{\mathrm{d} z}+\left(B^{2}-j(j+1)
            z(z+1)\right) R_{\omega j m}(r)=0,
        \end{eqnarray}
        where $B=\frac{(\omega-m \Omega_{eh})}{r_{e h}-r_{c h}} r_{e h}^{2}$.
        To get the equation above we used the approximations
        $\mathcal{Q} z \ll 1$ and $\mu_s^{2} r_{eh}^{2} \ll 1$, with $\mathcal{Q}=\left(r_{e h}-r_{c h}\right) \omega$.
         The latter
        comes from the consideration that the Compton wavelength
        of the scattered scalar field is much bigger than the size of the BH.
        The general solution of
        the above equation in terms of ordinary hypergeometric function ${ }_{2} F_{1}(a, b ; c ; z)$ is written as
        \begin{equation}
            R_{\omega j m}(z)=d\left(\frac{z}{z+1}\right)^{-iB}{
            }_{2} F_{1}\left(\frac{1-\sqrt{1+4j(j+1)}}{2},
            \frac{1+\sqrt{1+4 j(j+1)}}{2} ; 1-2 iB
            ;-z\right),\label{NEAR}
        \end{equation} where $d$ is a coefficient.
        Given that in the matching procedure we need to observe the
        large $z$ behavior of the above expression so the Eq. (\ref{NEAR})
        for case $z \to\infty$ turns into
        \begin{eqnarray}\label{NF}
            R_{\text {near-large } z} \sim d \left(\frac{\Gamma(\sqrt{1+4 j(j+1)})
                \Gamma(1-2 iB)}{\Gamma\left(\frac{1+\sqrt{1+4j(j+1)}}{2}
                -2 iB\right) \Gamma\left(\frac{1+\sqrt{1+4j(j+1)}}{2}\right)}
            z^{\frac{\sqrt{1+4j(j+1)}-1}{2}}+\right.\nonumber \\ \left.
             \frac{\Gamma(-\sqrt{1+4j(j+1)}) \Gamma(1-2
                iB)}{\Gamma\left(\frac{1-\sqrt{1+4j(j+1)}}{2}\right)
                \Gamma\left(\frac{1-\sqrt{1+4j(j+1)}}{2}-2 iB\right)}
            z^{-\frac{\sqrt{1+4j(j+1)}+1}{2}}\right).
        \end{eqnarray}

        Concerning the far-region, by taking approximations $z+1 \approx z$
        and $\mu_s^{2} r_{e h}^{2} \ll 1$ into account and dropping all the
        terms except those which describe the free motion with momentum $j$
        we get from the Eq. (\ref{RE2})
        \begin{equation}\label{FAR}
            \frac{\mathrm{d}^{2} R_{\omega j m}(z)}{\mathrm{d}
                z^{2}}+\frac{2}{z} \frac{\mathrm{d} R_{\omega j m}(z)}{\mathrm{d}
                z}+\left(k_{qh}^{2}-\frac{j(j+1)}{z^{2}}\right) R_{\omega j
                m}(z)=0~,
        \end{equation}
        where $k_{qh} \equiv \frac{\mathcal{Q}}{\omega}
        \sqrt{\omega^{2}-\mu_s^{2}}$. The
        general solution of Eq. (\ref{FAR}) is
        \begin{eqnarray}\nonumber
            R_{\omega j m, \text { far }}&=&e^{-i k_{qh} z}\left(f_{1} z^{\frac{\sqrt{1+4 j(j+1)}-1}{2}} M\left(\frac{1+\sqrt{1+4 j(j+1)}}{2},
            1+\sqrt{1+4j(j+1)}, 2 i k_{qh} z\right)+\right. \\
            &&\left.f_{2} z^{-\frac{\sqrt{1+4j(j+1)}+1}{2}}
            M\left(\frac{1-\sqrt{1+4j(j+1)}}{2}, 1-\sqrt{1+4j(j+1)}, 2 i k_{qh} z\right)\right), \label{FARR}
        \end{eqnarray}
where $M(a, b, z)$ is the confluent hypergeometric Kummer function
of the first kind. To match the solution above with
        (\ref{NF}), we have to find the small $z$ behavior of the solution
        (\ref{FARR}) which within the limit $z\to 0$ results in
        \begin{equation}
            R_{\omega j m, \text { far-small } \mathrm{z}} \sim f_{1}
            z^{\frac{\sqrt{1+4j(j+1)}-1}{2}}+f_{2}
            z^{-\frac{1+\sqrt{1+4j(j+1)}}{2}}. \label{FN}
        \end{equation}
        Now by matching of the two asymptotic  solutions (\ref{NF}) and (\ref{FN}) (as they
        have a common region of interest), we can determine coefficients $f_{1,2}$ as
        follow
        \begin{eqnarray}
            f_{1}=& d \frac{\Gamma(\sqrt{1+4j(j+1)})
                \Gamma(1-2 i B)}{\Gamma\left(\frac{1+\sqrt{1+4 j(j+1)}}{2}
                -2 iB\right) \Gamma\left(\frac{1+\sqrt{1+4j(j+1)}}{2}\right)}, \\\nonumber
            f_{2}=& d \frac{\Gamma(-\sqrt{1+4j(j+1)}) \Gamma(1-2
                iB)}{\Gamma\left(\frac{1-\sqrt{1+4j(j+1)}}{2}-2 iB\right)
                \Gamma\left(\frac{1-\sqrt{1+4j(j+1)}}{2}\right)}.
            \label{DD}
        \end{eqnarray}
        At this stage, by performing several consecutive  analytical steps   we will come to $\left|\mathcal{P}_{in}^{\infty}\right|$ and $\left|\mathcal{P}_{ref}^{\infty}\right|$,
 as two essential components involved in the extraction of scattering amplification factor $Z_{jm}$ or cross-section
        \begin{equation}
            Z_{j m} \equiv \frac{\left|\mathcal{P}_{r e
                    f}^{\infty}\right|^{2}}{\left|\mathcal{P}_{i
                    n}^{\infty}\right|^{2}}-1~. \label{AMPZ}
        \end{equation}
        Expanding the Eq. (\ref{FARR}) around infinity together with the
        approximations $\frac{1}{z} \sim \frac{\mathcal{Q}}{\omega} \cdot\frac{1}{r},
        \quad e^{\pm i k_{qh} z} \sim e^{\pm i\sqrt{(\omega^{2}-\mu_s^{2})} r}$
        and matching it with the radial solution \eqref{AS}, after inserting the
        expressions of $f_{1}$ and $f_{2}$ from Eq. (\ref{DD}),  we finally get
        \begin{eqnarray}
            \mathcal{P}_{in}^{\infty}&=&\frac{b(-2
                i)^{-\frac{1+\sqrt{1+4j(j+1)}}{2}}}{\sqrt{\omega^{2}-\mu_s^{2}}} \cdot
            \frac{\Gamma(\sqrt{1+4j(j+1)}) \Gamma(1+\sqrt{1+4j(j+1)})}{\Gamma\left(\frac{1+\sqrt{1+4j(j+1)}}{2}-2
                iB\right)\left(\Gamma\left(\frac{1+\sqrt{1+4j(j+1)}}{2}\right)\right)^{2}}\times
            \\\nonumber &&\Gamma(1-2 iB) k_{qh}^{\frac{1-\sqrt{1+4j(j+1)}}{2}}+\frac{b(-2
                i)^{\frac{\sqrt{1+4j(j+1)}-1}{2}}}{\sqrt{\omega^{2}-\mu_s^{2}}}
            \times \\\nonumber &&\frac{\Gamma(1-\sqrt{1+4j(j+1)})
                \Gamma(-\sqrt{1+4j(j+1)})}{\left(\Gamma\left(\frac{1-\sqrt{1+4j(j+1)}}{2}\right)\right)^{2}
                \Gamma\left(\frac{1-\sqrt{1+4j(j+1)}}{2}-2 iB\right)} \Gamma(1-2 i B)
            k_{qh}^{\frac{1+\sqrt{1+4j(j+1)}}{2}},
        \end{eqnarray}
        and
        \begin{eqnarray}
            \mathcal{P}_{ref}^{\infty}&=&\frac{b(2
                i)^{-\frac{1+\sqrt{1+4j(j+1)}}{2}}}{\sqrt{\omega^{2}-\mu_s^{2}}} \cdot
            \frac{\Gamma(\sqrt{1+4j(j+1)}) \Gamma(1+\sqrt{1+4j(j+1)})}{\Gamma\left(\frac{1+\sqrt{1+4j(j+1)}}{2}-2
                iB\right)\left(\Gamma\left(\frac{1+\sqrt{1+4j(j+1)}}{2}\right)\right)^{2}}\times
            \\\nonumber &&\Gamma(1-2 iB) k_{qh}^{\frac{1-\sqrt{1+4j(j+1)}}{2}}+\frac{b(2
                i)^{\frac{\sqrt{1+4j(j+1)}-1}{2}}}{\sqrt{\omega^{2}-\mu_s^{2}}}
            \times \\\nonumber &&\frac{\Gamma(1-\sqrt{1+4j(j+1)})
                \Gamma(-\sqrt{1+4j(j+1)})}{\left(\Gamma\left(\frac{1-\sqrt{1+4j(j+1)}}{2}\right)\right)^{2}
                \Gamma\left(\frac{1-\sqrt{1+4j(j+1)}}{2}-2 iB\right)} \Gamma(1-2 i B)
            k_{qh}^{\frac{1+\sqrt{1+4j(j+1)}}{2}}.
        \end{eqnarray}
 To find more details of the trend of the above calculations, we recommend referring to previous works \cite{Khodadi:2020cht, Khodadi:2021owg,Khodadi:2021mct} (also, review paper \cite{Brito:2015oca}).   Concerning Eq. (\ref{AMPZ}), there will be superradiance phenomena,
         precisely in the case
        $\frac{\left|\mathcal{P}_{ref}^{\infty}\right|^{2}}{\left|\mathcal{P}_{in}^{\infty}\right|^{2}}>1$
        i.e., when $Z_{jm} > 0$. Since we are interested in the
        occurrence of the superradiant phenomenon, until the end of this
        manuscript we ignored cases where $m \leq 0$ as they are
        non-superradiant. In what follows we will see how the hairy Horndeski parameter
         $h$ favorably affects the scalar superradiant scattering.

Through the display of the plots $Z_{11,22}-M\omega$ we evaluate
the role of Horndeski parameter $h$ on the amplification factor of
BH superradiance.  Fig. (\ref{hor}) clearly shows that in the
presence of hairy parameter the amplification factor of superradiance scattering and whose frequency range becomes larger and wider than the standard Kerr case, respectively. It means that the hairy Horndeski parameter
$h$ acts as an amplifier of the scalar wave and enhances the
chance of occurrence of  superradiance phenomena.

\begin{figure}[ht!]
    \begin{tabular}{c}
        \includegraphics[width=.5\linewidth]{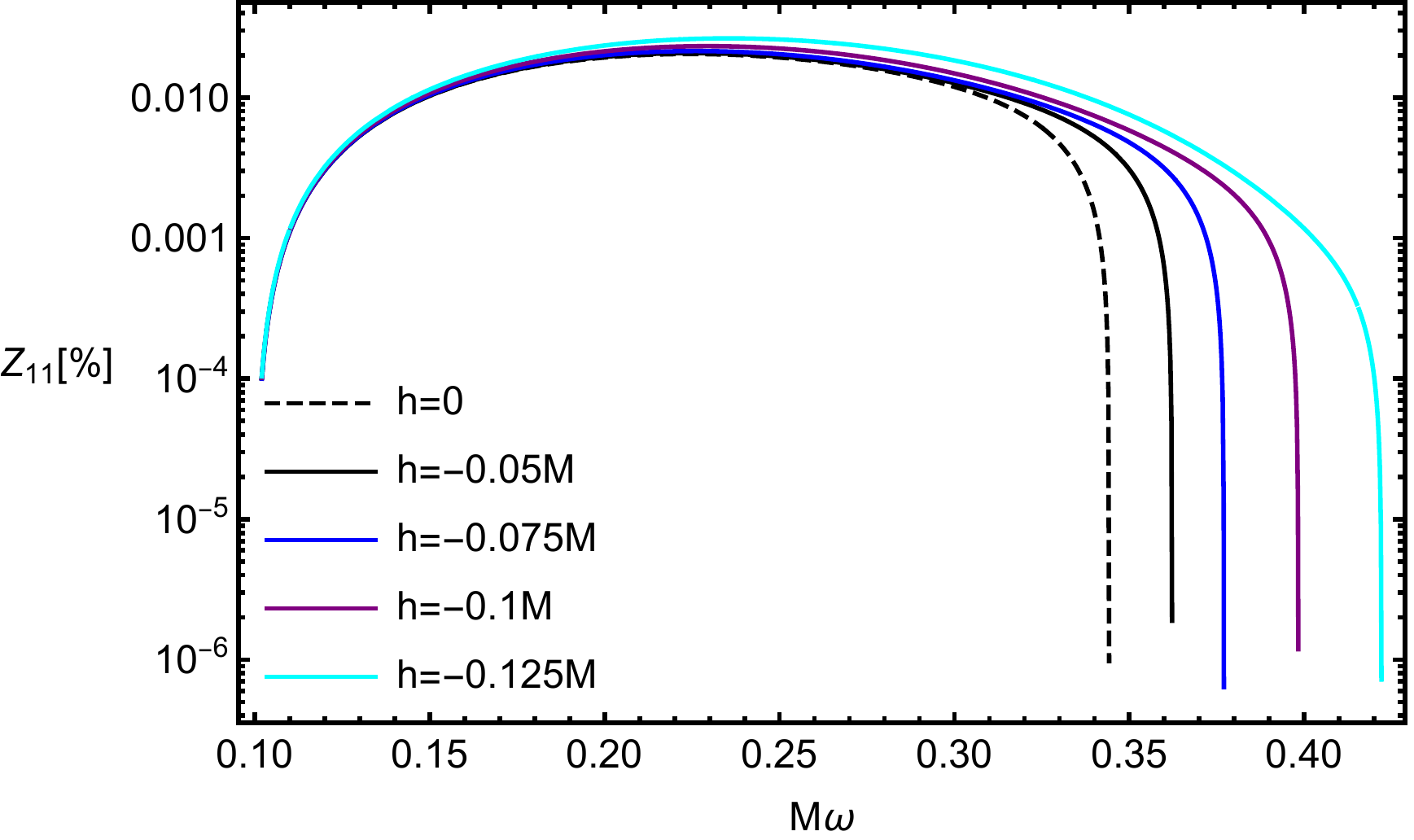}~~~
        \includegraphics[width=.5\linewidth]{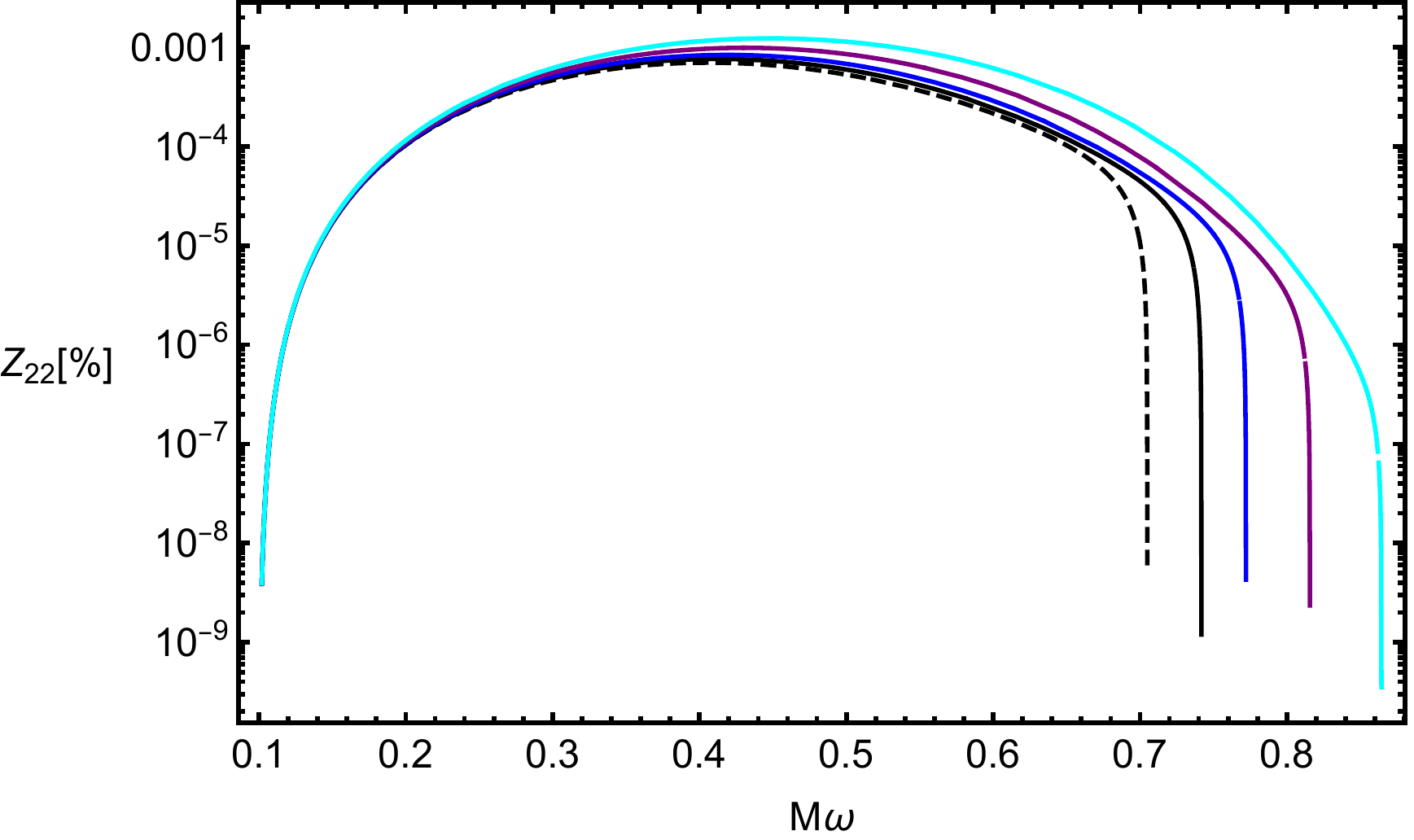}
    \end{tabular}
    \caption{\textit{Percentage amplification factors $Z_{11}$ and $Z_{22}$
    in terms of frequency for the rotating hairy Horndeski BH with variable
    values of hairy parameter $h$. Here and in the latter figures we take numerical
    values $\mu_s=0.1$ and $a=0.95M$ for the mass of the scalar bosonic field
    and the rotation parameter ratio of angular momentum to BH mass, respectively.
    Values fixed for $h$ in the right panel are the same left one.} } \label{hor}
\end{figure}

\begin{figure}[ht!]
    \begin{tabular}{c}
        \includegraphics[width=.5\linewidth]{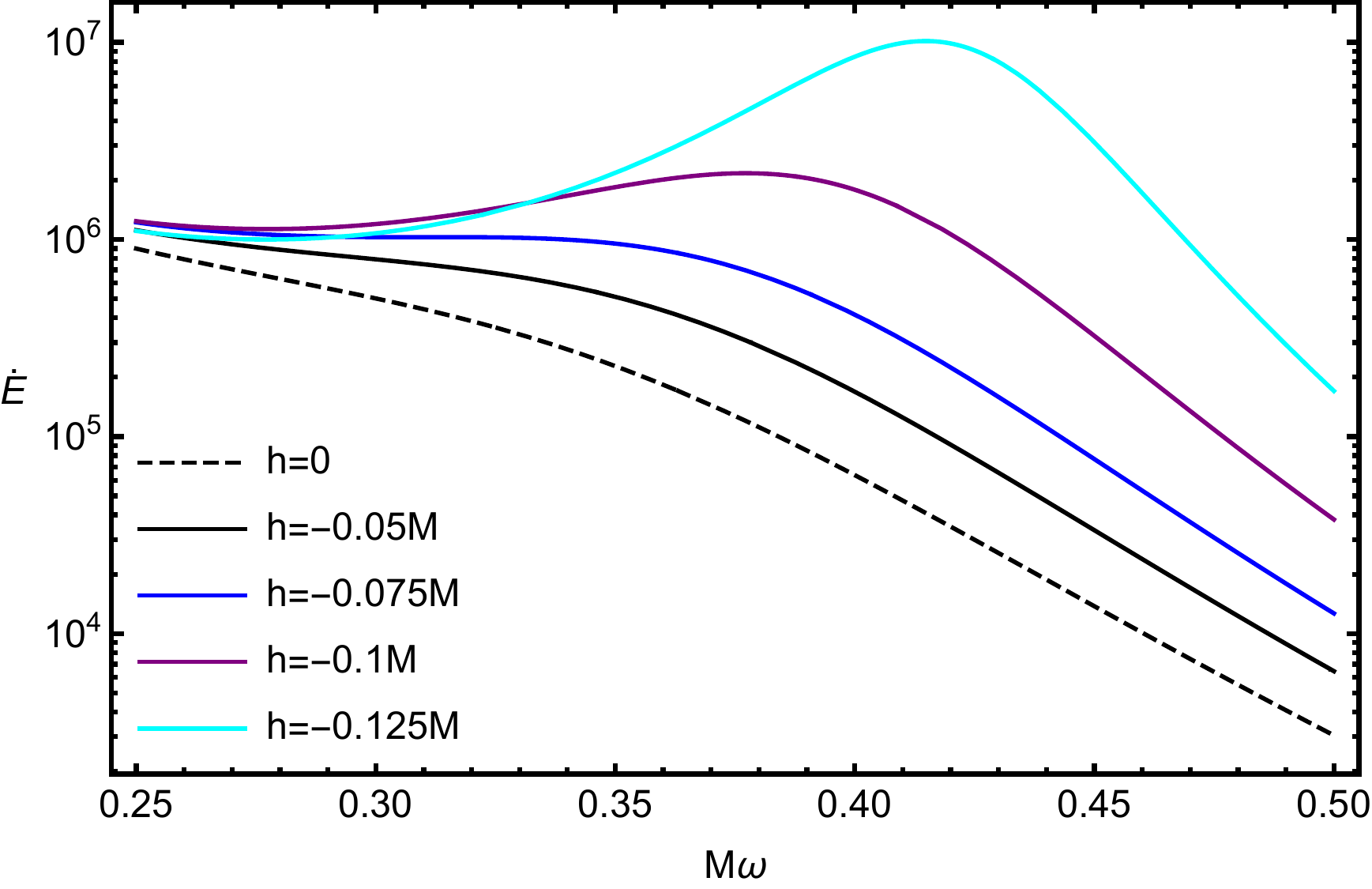}~~~
        \includegraphics[width=.5\linewidth]{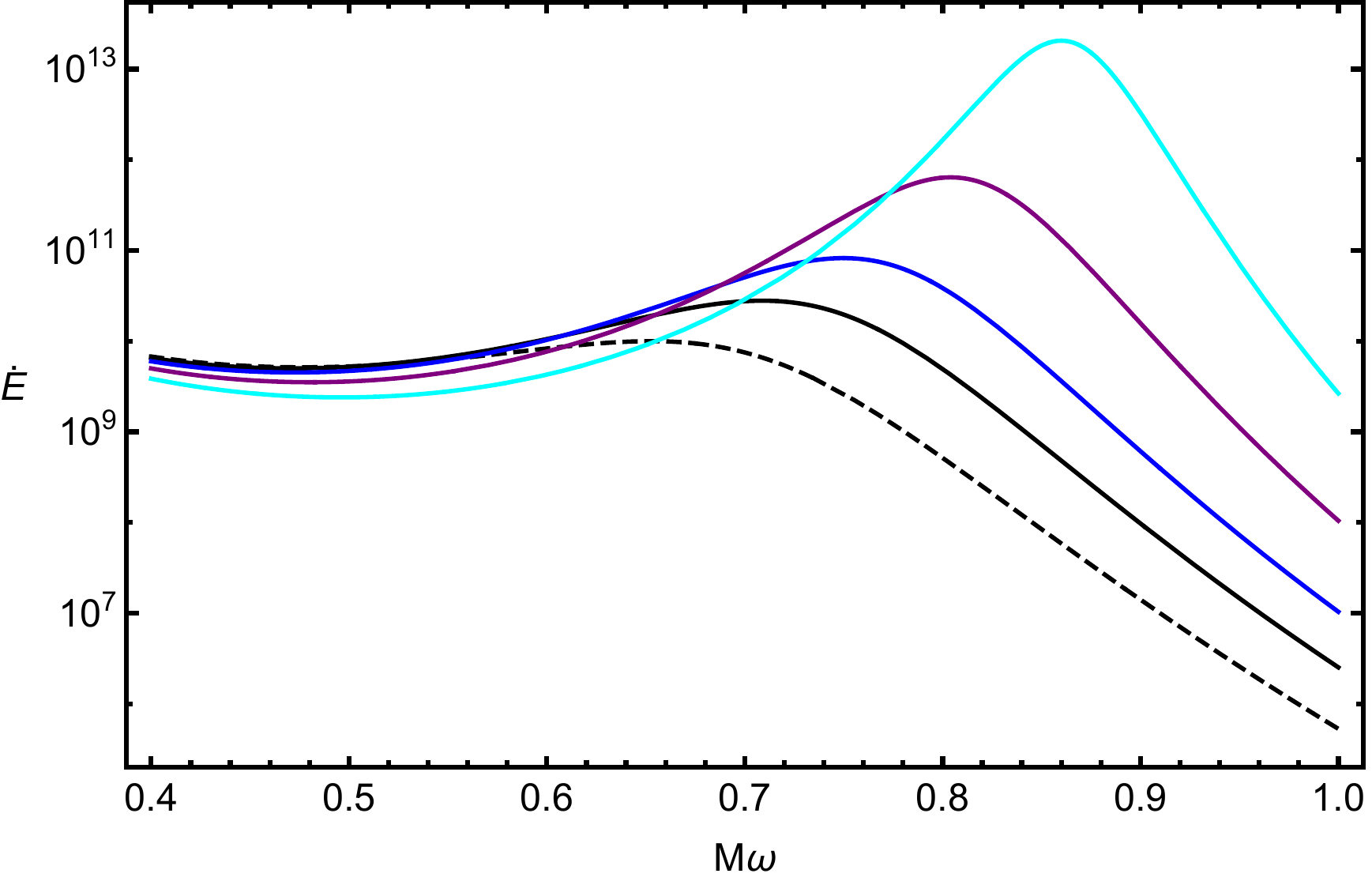}
    \end{tabular}
    \caption{\textit{Outgoing energy flux at infinity in terms of
    $M\omega$ for superradiant modes $l=1=m$ (left panel) and $l=2=m$
    (right panel). Values fixed for $h$ in the right panel are the same left one.} } \label{Lum}
\end{figure}
\subsection{Energy extraction}
Above revealed the role of hairy Horndeski parameter $h$ on
amplifying the scalar wave  scattered off a rotating hairy
Horndeski BH. This amplifying scenario is indeed equivalent to
energy extraction from BH. Here, we plan to show this via the
investigation of the impact of the hairy Horndeski parameter on
luminosity. The outgoing energy flux $\dot{E}$ measured by a
observer at infinity, namely for $r_{eh} \ll r \ll \mu_s^{-1}$,
can be calculated from the energy-momentum tensor of the field.
If the massive scalar field is monochromatic, one can derive it as
follows \cite{Brito:2015oca,Wondrak:2018fza}
\begin{eqnarray}
    \dot{E}=\frac{\omega k_{\infty}}{2} \left|\mathcal{P}_{ref}^{\infty}\right|^2 \left|\mathcal{P}_{in}^{\infty} \right|^2~.
    \label{lum}
\end{eqnarray}
By taking the superradiant modes into account of Eq. (\ref{lum}),
 we display plot $\dot{E}-M\omega$ in Fig. (\ref{Lum}).
One can see the effect of the enhancement of the extraction of
energy from the BH in comparison with the standard
  Kerr. Depending on setting different values for $h$ which ensure the
   nature of the compact object as BH, the energy extraction in peak
    frequencies that differ from
   case $h=0$ may boosted up to a few orders of magnitude.
    One effective technique to more clearly distinguish  Horndeski's and Einstein's theories
    of gravity is to integrate from Eq. (\ref{lum}),
    over the flux contribution of each mode weighted by the normalized
    initial mode distribution $n(\omega)$ i.e.,
\begin{eqnarray}
    \dot{E}_{tot}=\int d\omega \frac{\omega k_{\infty}}{2} \left|\mathcal{P}_{ref}^{\infty}\right|^2 \left|\mathcal{P}_{in}^{\infty} \right|^2 n(\omega)~.
    \label{lum1}
\end{eqnarray}
To determine the function related to $n(\omega)$, we consider a
thermal spectrum at  temperature $T$ for the incident spectrum
which obeys from
\begin{eqnarray}
n(\omega)=\frac{\omega^2}{2\zeta(3)k_BT^3\big(\exp[\omega/k_BT]-1\big)}~,
    \label{lum2}
\end{eqnarray}
where is the normalized black body mode spectrum
of a massless scalar field ($\mu_s=0$). Here, $\zeta(x)$ and $k_B$,
are the Riemann zeta function ($\zeta(3)\backsimeq1.2$), and Boltzmann
constant, respectively. By setting the temperature of the thermal spectrum
 around CMB ($\backsimeq2.7$K) as a background source, in Table. (\ref{Nu2}),
 we list values derived of numerically solving of Eq. (\ref{lum1}) in terms of
 the hairy parameter $h/M$ for superradiant modes at hand \footnote{The one-dimensional integral equation (\ref{lum1}) just like other analyses done in this paper is evaluated by the Mathematica software with version number (12.2.0.0) \cite{Wolfram}. An integration strategy prescribes how to manage and create new elements of a set of separate sub-regions of the initial integral region. Each sub-region might have its own integration rule related to it so that the integral estimation indeed is the sum of the integral estimates of all sub-regions. The integration rules to calculate the sub-region integral estimates, in essence, are created by sampling the integrand by a set of points so-called sampling points. To improve an integral estimate it should be sampled at additional points. There are two main approaches: adaptive and non-adaptive strategies. In the former is identifies the problematic integration areas and concentrate the computational efforts (i.e., sampling points) on them, while in the latter increase the number of sampling points over the whole region. The default strategy for numerical solving the integral in the Mathematica software i.e., an algorithm to compute integral estimates according to the precision specified by the user is so-called the ``Global Adaptive''. This method, in general, by recursive bisecting the sub-region with the largest error estimate into two halves reaches the required precision goal of the integral estimate and thereby, calculates the integral estimate for each half. }.
 The main message of the trend of numbers in the Table. (\ref{Nu2}),
 is that superradiant energy flux is amplified by parameter $h/M$, as shown earlier.

\begin{table}[ht!]
    \begin{center}
        \begin{tabular}{|c|c|c|}
            \hline
            $h/M$&$\dot{E}_{tot} (l=1=m)$&$\dot{E}_{tot} (l=2=m)$\\
            \hline
            $0$&$3.5\times10^4$&$3.65\times10^{7}$  \\  \hline
            $-0.05$&$7.3\times10^4$&$8.45\times10^{7}$ \\  \hline
            $-0.075$&$1.3\times10^5$&$2.01\times10^{8}$  \\ \hline
            $-0.1$&$4\times10^5$&$1.11\times10^{9}$  \\ \hline
            $-0.125$&$1.2\times10^6$&$2.13\times10^{10}$  \\ \hline
        \end{tabular}
        \caption{\textit{Numerical values of $\dot{E}_{tot}$ in terms of hairy parameter $h/M$ for a BH with rotation parameter $a=0.95M$, and superradiant modes: $l=1=m$ and $l=2=m$. }}
        \label{Nu2}
    \end{center}
\end{table}

\vspace{0.01cm}
\section{Superradiant instability regime} \label{Ins}
In this section, we are going to investigate the effects of the
hairy Horndeski parameter on the stability of rotating BH via a
phenomenon known as the BH bomb \cite{Press:1972zz}. The basic
idea behind this phenomenon is to use and enclose the extracted
rotational energy, by a mirror-like surface whether natural
(massive scalar field and AdS spacetime) or artificial (any
reflecting surface)- outside of BH for gradually growing and
amplifying waves via frequent round-trips. More technically, here
superradiant instability is a consequence of enclosing the massive
modes of a system composed of the Kerr background enriched with
hairy Horndeski parameter (\ref{rotmetric}) and the massive
scalar perturbations $\Phi$, inside the effective potential-well
placed outside the BH. In other words, to trigger superradiant
instability, the existence of a potential well outside the BH,
apart from the ergo-region, is essential.

Beginning from radial equation (\ref{RE}) we get
\begin{eqnarray}
\tilde{\Delta} \frac{d}{d r}\left(\tilde{\Delta} \frac{d R_{\omega
j m}}{dr}\right)+\mathcal{G} R_{\omega j m}=0, \label{MRE}
\end{eqnarray}
 where for a slowly rotating BH $(a \omega \ll 1)$
        $$
        \mathcal{G} \equiv \left(\left(r^2+a^{2}\right) \omega-ma\right)^{2}
        +\tilde{\Delta}\left(2 ma\omega-j(j+1)-\mu_s^{2}r^2\right).
        $$
        Following the BH bomb mechanism, we get the
        following solutions for the radial equation (\ref{MRE})
        $$
        R_{\omega j m} \sim\left\{\begin{array}{ll}
            e^{-i(\omega-m \Omega_{eh})} r_{*} & \text { as }~~~ r \rightarrow r_{e h}~~
            \left(r_{*} \rightarrow-\infty\right) \\
            \frac{e^{-\sqrt{\mu_s^{2}-\omega^{2} r_{*}}}}{r} & \text { as } ~~~r
            \rightarrow \infty~~~  \left(r_{*} \rightarrow \infty\right)
        \end{array}\right.
        $$
The above solution represents the physical boundary conditions
that the scalar wave at the BH horizon is purely ingoing while at
spatial infinity it is decaying exponential (bounded) solution,
provided that $\omega^{2}<\mu_s^{2}$. With the new radial function
        $$
        \psi_{\omega j m} \equiv \sqrt{\tilde{\Delta}} R_{\omega j m},
        $$
        the radial equation (\ref{MRE}) yields the Regge-Wheel equation
        $$
        \left(\frac{d^{2}}{d r^{2}}+\omega^{2}-V\right)
        \psi_{\omega j m}=0~,
        $$
        with

        $$
        \omega^{2}-V=\frac{\mathcal{G}+\mathcal{H}}{\tilde{\Delta}^{2}},
        $$
        where
        \begin{eqnarray}
     \mathcal{H}=  \frac{-2 a^2 (h+2 r)+r \left(h^2+2 h r+4 M^2\right)+h^2 r \log ^2\left(\frac{r}{2 M}\right)-4 h M r \log \left(\frac{r}{2
                M}\right)}{4 r}~.
        \end{eqnarray}
 %
        By discarding the terms
        $\mathcal{O}\left(1 / r^{2}\right)$, the asymptotic form of the
        effective potential $V(r)$ takes the following form
        \begin{eqnarray}\label{V}
             V(r)&=&\mu_s ^2-\frac{\left(2 \omega^2-\mu_s ^2\right) \left(2 M-h \log \left(\frac{r}{2 M}\right)\right)}{r}~.
        \end{eqnarray}
As a cross-check, one can see that if $h=0$, the above expression
recovers its standard form \cite{Hod:2012zza}.  The potential
represents trapping well when its asymptotic derivative is
positive i.e., $V^{\prime} \rightarrow 0^{+}$ as $r \rightarrow
\infty$ \cite{Hod:2012zza}. By taking derivative of potential
(\ref{V}), we have
\begin{eqnarray}\label{VV}
\frac{dV}{dr}=\frac{\left(2 \omega^2-\mu_s ^2\right)\bigg(2M+h(1-\log \left(\frac{r}{2 M}\right)\bigg)}{r^2}~.
\end{eqnarray}
Given that by default the hairy parameter is negative ($h < 0$),
then by demanding $\log \left(\frac{r}{2 M}\right)\geq1$ i.e.,
$r\gtrsim5.5M$, the expression above satisfies condition
$V^{\prime} \rightarrow 0^{+}$, if
  \begin{equation}\label{s}
\mu_s<\sqrt{2}\omega~,
\end{equation}
By taking this fact into account that the superradiance
amplification occurs when $\omega< m\Omega_{eh}$, so the
integrated system of Horndeski BH and massive scalar fields may
experience superradiant instability within the following regime
 \begin{equation}\label{ss}
    \mu_s<\sqrt{2}m\Omega_{eh}~,
 \end{equation}
  where is nothing but the standard superradiant instability regime is
  expected from a Kerr BH. Generally, for the
system at hand, the hairy Horndeski parameter $h$ has no
deterministic role in the superradiant instability regime.
\begin{table}[ht!]
	\begin{center}
		\begin{tabular}{|c|c|c|}
			\hline
			$h/M$&\mbox{Volume/M$^3$ ($a=0.9M$)}&\mbox{Volume/M$^3$ ($a=0.95M$)}\\	\hline
			$0$&$84.5$&$96.4$  \\  \hline
			$-0.05$&$89.3$&$104.8$ \\  \hline
			$-0.075$&$91.94$&$110.5$  \\ \hline
			$-0.1$&$94.83$&$118.45$  \\ \hline
			$-0.125$&$97.4$&$128.8$  \\ \hline
		\end{tabular}
		\caption{\textit{Numerical values of the ergoregion proper volume in terms of hairy parameter $h/M$ for a BH with  the rotation parameters: $a=0.9M$, and $a=0.95M$. }}
		\label{Nu3}
	\end{center}
\end{table}

\vspace{0.01cm}
\section{Closing remarks}\label{conc}
In this paper, we first took into account a spherically symmetric spacetime metric associated with the non-rotating Horndeski BH, and then to study the superradiance phenomenon we used the rotational version developed in \cite{Walia:2021emv}.
This metric is characterized by
 a hairy Horndeski parameter $h$, addressing a spacetime beyond Einstein's
 gravity. We have studied the superradiant scattering of the massive scalar test
 field in the background of the underlying spacetime and the extraction of
 energy from it. The motivation to apply such a modified geometry as a toy model for evaluating superradiant energy extraction comes from the fact that the statistical error reported in strong gravity tests potentially indicates small but detectable deviations from the standard Kerr BH.

Fig. (\ref{delta1}) reveals that depending on the suitable choice
for the hairy Horndeski parameter $h<0$ in interplay with spin
parameter $a$, the compact object arising from the metric at hand
is a BH. As has been seen, by fixing the value of $a$, for the
case of $h<0$, the spacing between the Cauchy horizon and event
horizon reduces, as the value of $h$ becomes more negative so that
after a certain value of $h$ the horizon disappears i.e.,
 there will be no longer a BH. Note that the critical value $h$ beyond which the spacetime can not be  BH  substantially depends on the rotation of the BH.

Given our interest in superradiant energy extraction from the BH,
until the end of our analysis, we have taken care of the
aforementioned point for choices of the values of $h$. It is clear
from Fig. (\ref{hor}) that for the rotating Horndeski BH, if the
value of the hairy parameter $h$ becomes more negative, thereby,
the superradiance amplification factor and whose frequency range
rises and widens respectively relative to the standard Kerr
case that corresponds to case $h=0$.
 It means that in an extended framework of gravity as the Horndeski model,
 scalar wave-based superradiance scattering is strengthened. In this direction,
 by deriving outgoing energy flux for the faraway observer, we have
 addressed explicitly the role of hairy parameter $h$ on energy extraction
  from BH. In agreement with the enhancement of the superradiance
  amplification factor by hairy parameter $h$, we have demonstrated
that  it causes the increase of energy extraction from the
rotation BH. It  can be easily verified through Fig. (\ref{Lum}) and Table. (\ref{Nu2}). As a consequence, the hairy parameter $h$ amplifies the scalar wave and enhances the chance of the superradiance. Adding some comments here to understand the physics behind this phenomenon can be helpful.  Recall that the friction and some negative-energy states for energy extraction via superradiance are essential. Concerning rotating BHs both, in essence, are supplied by the ergoregion as a region near the event horizon in which the energy of timelike particles is negative \cite{Brito:2015oca}. Besides, it is well-known that the background geometry plays role in increasing/decreasing the amplification factor of the wave so that the strengthening/weakening of the scattered waves from the rotating BHs, understand in terms of the increase/decrease of the proper volume of the ergoregion. Namely, between the ergoregion proper volume and the superradiant amplification factor, there is a correlation, meaning that the larger the former, the more time the wave spends in the ergoregion and the more energy it extracts from BH \cite{Brito:2012gj}. 
As a result, the modifications induced on the background geometry by alternative theories of gravity affect the proper volume, and the amplification factor subsequently. More precisely, as the proper volume of the ergoregion increases/decreases, more/less energy compared to Kerr is extracted through superradiance scattering. To support this statement, it is enough one calculate the the proper volume via $V=4\pi \int_{\theta_i}^{\pi/2}d\theta\int_{r_i}^{r_f} dr\sqrt{g_{rr}g_{\theta\theta}g_{\phi\phi}}$ which for case of $a<M$ ergoregion extends from $r_i$ to $r_f$ i.e., between the location of event horizon and outer ergosphere radius \cite{Pani:2010jz}. In the Table. (\ref{Nu3}), by setting $\theta_i=0$ for two cases $a=0.9M$, and $0.95M$, we release the numerical values of the proper volume of ergoregion in terms of different values of hairy parameter $h$.
As can be seen in the presence of the hairy parameter $h$ correction, the proper volume of the ergoregion increases compared to standard Kerr BH, meaning that it behaves as an amplifier of the scalar wave.

The importance and worth of these results become especially
clear when we contrast them with earlier findings from
scalar-tensor theories utilizing Kerr BH surrounded by the matter
profile \cite{Cardoso:2013opa,Cardoso:2013fwa}. It was demonstrated that the amplification factor in scalar-tensor gravity can be higher than in the typical situation because of scalar-matter interactions. For Horndeski gravity, as the most general four-dimensional scalar-tensor theory, this enhancement in the amplification factor occurs even in the absence of enclosing the BH by a matter profile.

By analyzing the effective potential within the context of BH bomb
mechanism, we have surveyed the superradiant instability of the
rotating Horndeski BH, subjected to massive scalar perturbation.
We found out that the hairy Horndeski parameter $h$ leaves no
effect on the standard superradiant instability regime.
\\\\

\vspace{0.01cm}
\textbf{Acknowledgments:}
We appreciate the anonymous referee for insightful comments that helped us improve the paper.

    \end{document}